\documentclass[pra,twocolumn,amsmath,amssymb,groupaddress,nofootinbib]{revtex4-2}

\usepackage{graphicx,relsize,epstopdf,color,mathtools,bm,newtxtext,newtxmath,braket,rotating,dsfont,mathtools}

\usepackage[hyphenbreaks]{breakurl}
\usepackage[colorlinks=true,linkcolor=blue,citecolor=blue,urlcolor =blue]{hyperref}
\usepackage{float}
\usepackage{flafter}
\usepackage{placeins}
\usepackage{soul}
\usepackage{cancel}
\usepackage{comment}

\begin{document}

%\preprint{APS/123-QED}

\title{Light and divergences:  History and outlook}

\author{Gerd Leuchs}
\affiliation{Physik Department, Naturwissenschaftliche Fakultät, Friedrich-Alexander-Universität Erlangen-Nürnberg, 91058 Erlangen, Germany} 
\affiliation{Max-Planck-Institut f\"{u}r die Physik des Lichts, 91058 Erlangen, Germany}
\affiliation{Department of Physics, Faculty of Science, University of Ottawa, Ottawa ON K1N7N9, Canada}
\author{Luis L. S\'anchez-Soto}%
\affiliation{Departamento de \'Optica, Facultad de F\'{\i}sica, Universidad Complutense, 28040 Madrid, Spain}
\affiliation{Max-Planck-Institut f\"{u}r die Physik des Lichts, 91058 Erlangen, Germany}

\date{\today} 

\begin{abstract}
All experimental evidence {indicates}  that the vacuum is not void, but filled with something truly quantum. This is reflected by terms such as {zero-point} fluctuations, and Dirac's sea of virtual particle-antiparticle pairs, and last but not least the vacuum is the medium responsible for Maxwell's displacement current. While quantum electrodynamics (QED) is an exceptionally successful theory, it remains a perturbative framework rather than a fully self-contained one. Inherently, it includes singularities and divergences, which prevent the precise calculation of fundamental quantities such as the fine-structure constant $\alpha$. Any direct attempt to compute $\alpha$ results in divergent values. However, and most remarkable, what can be determined is how $\alpha$ ``runs", meaning how it varies with energy or exchanged momentum. In this article, we review the historical development of these ideas, the current state of knowledge, and ongoing efforts to find ways to move further. This includes a simple model to describe vacuum polarization in the low-energy regime, when considering only small (linear) deviations from the equilibrium {state}, relating  {Maxwell's displacement} in the vacuum, to the quantum properties of the vacuum.\footnote{We dedicate this article to Joe H. Eberly who passed away unexpectedly on 30 April 2025 short before his 90th birthday. He was a constant source of inspiration for us over several decades until earlier this year. We will fondly remember our discussions with him and we will miss him.}
\end{abstract}

\maketitle

%\tableofcontents

\section{\label{sec:level1}Introduction}

The way physicists perceive nature has undergone a profound transformation over the past 125 years. Remarkably, fundamental questions remain unanswered—and most likely, new ones will continue to emerge, ensuring that physicists will never run out of work!  In this article, we take a non-standard perspective on quantum electrodynamics (QED). QED {holds a unique place in physics: though plagued by troublesome divergences, it has nonetheless achieved some of the most precise predictions in all of science.}   So far, it remains a perturbative theory. though, requiring ever-higher orders of perturbation to achieve greater accuracy. But {So how did this remarkable journey begin?}    

In the second half of the nineteenth century, the microscopic approach to thermodynamics emerged, leading to numerous breakthroughs. One of Ludwig Boltzmann's key discoveries was that the average thermal energy per degree of freedom is $kT/2$, where $k$ is Boltzmann’s constant and $T$ the temperature. By the early 1900s, one of the prominent unsolved problems was the spectrum of blackbody radiation~\cite{Kuhn:1978aa,Boyer:2018aa}. A straightforward application of the newly developed field of thermodynamics to this problem required first to calculate the number of standing waves in a volume $V$ with frequencies in the interval from $\nu$ to $\nu+d\nu$ and then multiply it by the mean thermal energy per standing wave.
Since each standing wave possesses two degrees of freedom, much like a single harmonic oscillator, the mean energy is $kT$. The expression in modern notation can be written as~\cite{Rayleigh:1900aa,Jeans:1905aa}:
\begin{equation}
\rho (\nu) = \frac{8 \pi \nu^2}{c^3} kT \, ,
\label{RayleighJeans}
\end{equation}
with $\rho (\nu )$ being the radiation energy per volume and per frequency interval $d\nu$ {and $c$} the speed of light in vacuum.  Integrating over all frequencies results in an infinite total energy, a problem referred to as the ultraviolet catastrophe~\cite{Ehrenfest:1911aa}, since $\rho (\nu )$ diverges as $\nu$ approaches infinity. This was the first time a divergence appeared. 

The mean energy $\langle w\rangle =kT$ can be calculated using Boltzmann’s distribution {$p(\omega) \propto \exp(- w/kT)$}   and $w$ as a continuously variable quantity. But, at about the same time, Max Planck introduced a crucial modification. He postulated that the energy exchange between matter and light occurs in discrete units, or quanta, of $h\nu$, where $h$ is a fundamental constant that now bears his name.   With this assumption, the calculation of the mean energy $\langle w_n \rangle= { n h} \langle \nu \rangle$ has to be done using a sum and not an integral. This eliminated the divergence and led to Planck’s first famous radiation formula~\cite{Planck:1901aa}
\begin{equation}
    \rho (\nu )=\frac{8\pi\nu^2}{c^3}\ \ \frac{h\nu}{e^\frac{h\nu}{kT}-1} ,	
\label{planck1}
\end{equation}
which matched experimental observations remarkably well across a wide spectral range. These experiments, conducted around the same time, were most challenging. Two of the key experimental pioneers were Otto Lummer and Ernst Pringsheim \cite{Lummer:1900aa}. They played a crucial role in obtaining precise blackbody radiation data.

To extend measurements far into the infrared, new spectrally resolving detectors had to be developed. Heinrich Rubens, another pioneer in this field, leveraged the \emph{Reststrahlen} effect~\cite{Rubens:1900aa} to enhance infrared detection, enabling convincing experimental confirmation of Planck’s theoretical predictions.

All seemed fine, but then Planck noticed that his original formula \eqref{planck1} did not match the Rayleigh-Jeans law \eqref{RayleighJeans} 
perfectly in the high temperature limit (as it was expected to do) but that there was a small temperature-independent offset. So in 1912, he modified his first formula by adding $\frac{1}{2}h\nu$ to the energy per mode, recovering the convergence \cite{Planck:1912aa}. And the catastrophe came back: the energy density $\rho (\nu )$ now again diverges to infinity as the frequency increases. The paper describes his struggle with this problem. Planck himself acknowledges the limited applicability of this temperature-independent term in the energy per mode, as it results in a total energy diverging in any finite volume. He referred to it as ``latent energy" and engages in bold speculation about its implications. This is Planck’s second radiation formula:
\begin{equation}
    \rho (\nu) =\frac{8\pi\nu^2}{c^3}\ \ \left(\frac{h\nu}{e^\frac{h\nu}{kT}-1}+\frac{1}{2}h\nu\right) \, .
\label{planck2}
\end{equation}
This divergence has persisted to the present day and is closely related to the divergences that emerged later during the development of QED towards the end of the 1930's. 

Soon after Planck had published~ Eq.\eqref{planck2}, there came its first success: using this formula, Einstein and Stern~\cite{Einstein:1913aa} succeeded in explaining the specific heat near $T=0 $ for certain molecular gases, a phenomenon previously unexplainable. 
Thus, the ground-state energy $\tfrac{1}{2}h\nu$ of a harmonic oscillator appeared in the literature years before Heisenberg, Schrödinger, and Dirac formulated their respective  equations. A decade ago, a comprehensive book was published, discussing and speculating on the consequences of Planck's second blackbody radiation formula~\cite{Kragh:2014aa}.

Although the ground state energy is finite, it cannot be extracted or utilized, hence Planck referred to it as being latent. However, it does lead to measurable effects, such as spontaneous emission, the Lamb shift, and the Casimir effect~\cite{Milonni:1994aa}.  Although Planck had used thermodynamic reasoning to argue for the finite ground-state energy, Dirac introduced a second quantization, which formally assigned a ground-state energy of $\frac{1}{2}h\nu$ to each mode. Since then, field theories have grappled with the divergences associated with this energy, even after Schwinger, Feynman, Tomonaga and Dyson introduced renormalization~\cite{Itzykson:1980aa,Bogoliubov:1980aa,Peskin:1995aa,Weinberg:1995aa}.

Before Einstein, physicists believed that the vacuum was not void, but contained a medium necessary for light propagation, which they called ether or \ae ther. The concept gained clarity when Maxwell realized that a self-consistent theory of electromagnetism required that applying an electric field to the vacuum induces a displacement current, analogous to what happens in a dielectric.  In SI units the displacement is written as $D=\varepsilon_{0}  \chi_{\mathrm{vac}}E+P=\varepsilon_{0} E+P$, the displacement current of the vacuum being the temporal derivative of the second term: $\dot D_{\mathrm{vac}}=\varepsilon_{0}\dot E$. Could it be that Maxwell’s displacement current and the ground-state energy per mode are connected? After all, both occupy the vacuum. In fact, it seems that the so-called vacuum fluctuations of virtual elementary particle-antiparticle pairs, the zero-point fluctuations of the electromagnetic field associated with the ground-state energy per mode, and Maxwell’s displacement might merely be different sides of the same coin.

\section{\label{sec:level1}  Setting the problem}

How big a challenge it was to formulate a new theory of electrodynamics {that incorporated} quantum physics is underlined by {by the time and effort it demanded.} {Throughout the 1930's, the problem occupied an entire generation of physicists, yet it took more than a decade after Dirac’s groundbreaking hypothesis of antimatter before a workable framework emerged. It was only in the mid-1940's that}   Feynman, Schwinger and Tomonaga came up with the formulation of QED that is still used today.

The approach that was finally successful was to axiomatically postulate Maxwell's equations with the electric and magnetic fields as operators according to the second quantization describing light. {Simultaneously, Dirac's equation was adopted to govern the behavior of} the electron. The interaction between the electron and the {quantized} field was then explicitly introduced~\cite{Mehra:2000aa}.

Therefore, certain aspects of the quantum vacuum are axiomatically incorporated  in Maxwell’s equations, while others are addressed through the explicit interaction between matter and the field. On the one hand, Maxwell’s equations inherently account for the linear interaction of the electromagnetic field with the vacuum{—an essential backdrop without which the very concept of a field loses meaning.} Indeed, there is no electromagnetic field without the vacuum~\cite{Zeldovich:1967aa} and the speed of electromagnetic waves in vacuum results directly from these underlying principles~\cite{Meis:2017aa}. On the other hand, any nonlinear interaction between light and the vacuum, as well as all interactions involving real matter, are incorporated through explicitly added interaction terms.

The interaction between light and matter is governed by a coupling constant, which in QED can be defined in multiple ways. The most common choice is Sommerfeld's fine structure constant $\alpha$, because it is a dimensionless number. Alternatively, one can choose the square of the elementary charge $e^2$. {A further complication arises from the fact that the vacuum can itself be treated as a dielectric medium (as discussed below)}, a concept intimately related to the displacement current that Maxwell introduced to ensure the self-consistency of electrodynamics.  This also results in the screening of the charge of pointlike elementary particles.  During an electron-electron collision, their effective charges appear to increase with increasing momentum exchange, as each electron penetrates the other's screening cloud. This phenomenon is referred to as the running of the coupling. 
At low momentum exchange, the QED coupling constant is $\alpha (0) = {e^2}/{(4\pi \varepsilon_0 \hbar c_{\mathrm{rel}})}$. This is the expression in SI units and $c_{\mathrm{rel}}$ is the limiting speed in Lorentz's transformation. To ensure that we capture all relevant physical effects, we should express it as follows: 
\begin{eqnarray} 
\alpha (0) = \frac{e^2}{4\pi \varepsilon_0 \chi_{\mathrm{vac}}(0) \hbar c_{\mathrm{rel}}}. 
\label{alphachivac} 
\end{eqnarray}
$\chi_{\mathrm{vac}} (0)= 1$ is the susceptibility of the vacuum, which is unity everywhere without dispersion as a result of the Lorentz invariance and is often omitted in equations. However, we retain it because we want to find an expression for the running as a function of the exchanged momentum $p^2 = \hbar^2 k^2$ through the connection to Maxwell \footnote{Note that we use the reciprocal $k$ space and since this response must be Lorentz invariant, it must be a function of $k^{2} = \omega^{2}/c_{\mathrm{rel}}^{2}-\mathbf{k}^{2}$.  The condition $k^{2}=0$, describing a freely propagating photon, is referred to as on-shellness in QED: a real on-shell photon  verifies then $\omega^{2}=\mathbf{k}^{2}c_{\mathrm{rel}}^{2}$. However, in collisions and other situations where one has nonpropagating fields, such as evanescent waves or near fields, $k^{2}$ will typically not be zero.}. 

In Gaussian units $\varepsilon_0$ becomes unity, much like $\chi_{\mathrm{vac}}(0)$. What counts physically is their product.  When calculating $\chi_{\mathrm{vac}} (k^2)$ and thus $\alpha (k^2)$ through vacuum polarization in standard QED to second order in perturbation theory, the result diverges instead of yielding $1$ for $k^2 = 0$. This is one of the fundamental divergences one encounters. As a result, we learn that we cannot calculate $\alpha (0)$ and that we can only calculate how it runs.

Considering all different types of  virtual elementary particles with electrical charge $q_i$ and mass $m_i$, second order perturbation theory yields \cite{Itzykson:1980aa}:
\begin{eqnarray} 
\alpha^{-1} (k^{2}) &  = & \alpha^{-1} (0) - \frac{1}{3\pi} \sum_{i}^{\mathrm{e. \, p.}} \frac{q_i^2}{e^2} \int_{0}^{1} d x \, x (1-x) \nonumber \\
& \times & \ln \left[  1  + x(1-x) \frac{\hbar^{2} k^{2}}{m_i^{2}c_{\mathrm{rel}}^{2}} \right ] \, ,
\label{running alpha} 
\end{eqnarray}
where the sum is over all different types of elementary particles. 

As the momentum exchange increases, the screening effect is gradually lost: first due to the least massive particles and then due to the more massive ones.  These changes roughly happen when $\hbar^2 k^2 = m_i^2 c_{\mathrm{rel}}^2$. The resulting  running~\cite{Hogan:2000aa}, i. e. the $k^{2}$-dependence, was tested experimentally in a momentum exchange regime where the difference between $\alpha (k^2)$ and $\alpha (0)$ amounted to a few percent \cite{Abbiendi:2004aa}. At much higher momenta one of course expects higher orders of perturbation theory to dominate, though this does {not,} which does not necessarily eliminate the divergence. 

One can, however, use equation \eqref{running alpha} for the running $\alpha (k^2)$ to come up with a closed mathematical expression for $\alpha (0)$. 
For sufficiently large $k^2$, $\alpha^{-1}(k^2)$ will eventually go to zero at the hypothetical momentum $\hbar^2 k^2 = \Lambda^2$. As a tribute to Lev Landau, this is referred to as the Landau pole~\cite{Landau:1954aa} and represented by the symbol $\Lambda$. At such high values of  $k^2$, the equation for the running of $\alpha$ can be asymptotically simplified:
\begin{eqnarray} 
\alpha^{-1} (0) = \frac{1}{3\pi} \sum_{i}^{\mathrm{e. \, p.}} \frac{q_i^2}{e^2} \ln \left (  \frac{\Lambda^2}{m_i^{2}c_{\mathrm{rel}}^{2}} \right ) .  
\label{alpha large k2} 
\end{eqnarray}
But, of course, this only shifts our ignorance from not knowing $\alpha (0)$ to not knowing $\Lambda$. Already in 1967, Zel'dovich came up with a very similar expression for the coupling \cite{Zeldovich:1967aa}, which, if adapted to our notation reads as:
\begin{eqnarray} 
\alpha^{-1} (0) = \frac{1}{3\pi} \nu \ln \left (  \frac{\Lambda^2}{m^{2}c_{\mathrm{rel}}^{2}} \right ) .  
\label{Zeldovichi} 
\end{eqnarray}
Zel'dovich wrote this paper short before the discovery of quarks. At that time, it was assumed that all charged elementary particles had the same elementary charge $e$. In Zel'dovich's formula the coupling constant is likewise given by a cutoff at the Landau pole and the number {$\nu$ refers here to the } different types of elementary particles (compare to the sum over squared charges in the formula above!). Both the former, $\Lambda$,  and the latter, $\nu$, we cannot be sure about. Maybe we have not yet discovered all elementary particles, mind you that there are speculation about super-symmetry. And the value for the Landau pole we have to guess too. Zel'dovich speculated that the Landau pole momentum is the Planck momentum $\Lambda^2 = {\hbar c_{\mathrm{rel}}^3}/{G}$, where $G$ is the gravitational constant~\cite{Sakharov:1991aa}. These are all interesting potential links, but QED does leave some questions unanswered. We can only speculate that a rigorous quantum field theory --if it could be found-- would be able to relate the fine structure constant to all the different types of electrically charged elementary particles, already known or not yet known by mankind. On the other hand, it is most remarkable, though, how many predictions of QED match extremely well with experimental results, despite of the non-physical divergences, which are most likely a result of the perturbative nature of the theory.

\section{Marching off the trodden path}

Following Dirac's hypothesis, several physicists proposed that the quantum vacuum behaves like a dielectric medium (see, e. g., \cite{Furry:1934aa,Pauli:1934aa,Weisskopf:1936aa,Dicke:1957aa,Gottfried:1986aa}). In view of the preceding discussion, this analogy suggests the possibility of calculating the vacuum permittivity and thus the QED coupling constant $\alpha$. Several groups started to embark on this road, assuming that the response of the unit cell in the vacuum; e. g., the one associated to a single virtual electron-positron pair, can be modeled using a harmonic oscillator~\cite{Leuchs:2010aa,Mainland:2020aa}. The volume of this unit cell can be roughly estimated using Heisenberg's uncertainty relation~ \cite{Leuchs:2013aa,Urban:2013aa,Mainland:2019aa,Leuchs:2023aa}. There are numerous examples in physics, where the dynamics of small deviations from a thermal equilibrium of a system can be well described by a harmonic analysis. Moderate light powers certainly introduce only minute excursions from the vacuum in equilibrium. Thus one might expect that such an approach could work well. Such paths were followed by  ~\cite{Leuchs:2010aa,Mainland:2020aa} in different ways. The former, we will sketch in the following. 

Our understanding of the physical properties of the vacuum lies at the core of the discussion surrounding the divergences we encounter. It is fascinating to recognize, in retrospect, that Maxwell's seminal work--suggesting a dielectric nature of the vacuum--is deeply connected to quantum physics. This realization motivated us to explore the link between Maxwell’s framework and quantum theory.

The starting point was to investigate whether the electric and magnetic susceptibilities of the vacuum could be derived from our current understanding of the quantum vacuum. A key challenge is that these susceptibilities are not frequency-dependent. In Gaussian units, they are normalized to unity, whereas in SI units, they manifest as the dimensional constants $\varepsilon_0$ and $\mu_0$.
Attempting to compute these two fundamental quantities in an open-ended manner presents an additional complication: when doing so, one must pretend that the speed of light is not inherently known. But the knowledge of the limiting speed of relativity is needed in the approach as we will see below. Therefore, we have denoted by $c_{\mathrm{rel}}$ the speed that relates the mass of a particle $m$ to its rest-energy $E=mc_{\mathrm{rel}}^2$. Whether or not the speed of light $1/\sqrt{\varepsilon_0 \mu_0}$ resulting from the calculation of $\varepsilon_0$ and $\mu_0$ matches with  this limiting speed $c_{\mathrm{rel}}$ or not will then be a stringent test to check the validity of the model. Along this line,  {we can write}
\begin{eqnarray} 
\mathbf{D}(\mathbf{r}, t) = \mathbf{P}_{\mathrm{vac}} (\mathbf{r}, t) + \mathbf{P}_{\mathrm{mat}} (\mathbf{r}, t)  = 
\varepsilon_0 \, \mathbf{E} (\mathbf{r}, t) + \mathbf{P}_{\mathrm{mat}}(\mathbf{r}, t) .
\label{Displacement} 
\end{eqnarray}

As mentioned earlier, the notion that the quantum vacuum behaves like a dielectric has been repeatedly discussed in the literature. However, there is a fundamental difference compared to ordinary dielectric materials. In conventional dielectrics, the unit cell volume or the number of molecules per unit volume is well defined. In contrast, for the quantum vacuum, it is not immediately clear what volume should be assigned to a single virtual particle–antiparticle pair. There are several options: 

\begin{enumerate}
    \item Motivated by Heisenberg's uncertainty relation one might guess that it is the cube of the Compton wavelength of the respective type of elementary particles: 
    \begin{equation}
    V_{\mathrm{pair}} = \frac{\hbar^3}{m^3c_{\mathrm{rel}}^3}\, .
    \end{equation}  
    Variations of this argument were used  in 
    \cite{Leuchs:2013aa,Urban:2013aa,Mainland:2019aa}.
    \item Alternatively, one could ask how many pairs fit into a given volume? and calculate the number of standing particle waves of wavelength $\lambda = {h}/{p}$ in the volume, much like in the related calculation for light {[see \eqref{RayleighJeans}]}. In this case, the spectral density increases with $p^2$ and the total number diverges to infinity --in turn one would assign zero volume to a single virtual pair: $V_{\mathrm{pair}} = 0$. Dividing by a zero volume will obviously lead to a divergence.
    
    \item Not satisfied with this divergence, we might decide to modify approach (2) {by introducing}  a relativistic cutoff at roughly $p = 2mc_{\mathrm{rel}}$, $2m$ being the mass of a {virtual} pair. {This kind of \textit{cutoff} was among the techniques employed during the early development of QED to regularize divergent integrals and yields}  
    \begin{equation}
    V_{\mathrm{pair}} = \frac{3\pi^2}{4} \frac{\hbar^3}{m^3 c_{\mathrm{rel}}^3} .
    \end{equation}
    \item  Or, we let quantum physics may come to the rescue. In this approach, we recall that the goal is to determine the response of the quantum vacuum dielectric to an external electric field. In the absence of a field, the vacuum remains in equilibrium; with an applied field, a small deviation from equilibrium occurs. Since small perturbations around equilibrium are often well-described by a harmonic response, we model each virtual particle-antiparticle pair as a quantum harmonic oscillator. In this framework, two parameters fully define the system: the mass and the frequency of the oscillator. Since the frequency is related to the energy gap between the ground state and the lowest excited state: $\omega = 2mc_{\mathrm{rel}}^2/{\hbar}$, 
    the dynamics of the harmonic oscillator is determined by just a single parameter: the mass of the particle. This sole parameter also determines  
    the spatial extension of the ground state wavefunction\footnote{{It} one applies the same harmonic model to the hydrogen atom, using the energy gap between the ground and the first excited state to determine $\hbar\omega$, the root mean square variance for the position variable in the ground state is off by only 15\% when comparing to Bohr's radius! The result also compares favorably well (within 30\%) with the interatomic distance in solid state hydrogen \cite{McMahon2011}.}: 
    \begin{equation}
    V_{\mathrm{pair}} = \left ( \frac{\pi}{4} \right )^\frac{3}{2}  \frac{\hbar^3}{m^3c_{\mathrm{rel}}^3} .
    \end{equation}
    To obtain this expression, we approximated the modulus squared  of the Gaussian ground-state wavefunction using an equivalent rectangular distribution of equal height and integrated area. Additionally, we introduced a factor $2^{-3}$ to account for the spatial extension of the actual motions of the particle-antiparticle pairs being a factor $2$ smaller in each of the three spatial dimensions in comparison to the equivalent harmonic model with a single particle having the reduced mass $m/2$. While this is still an approximate model it seems to be the one with least room for adjustment so far for the resulting volume per pair.
    
    \item The fifth and final option discussed here is also based on the quantum harmonic oscillator wavefunction. However, instead of approximating the modulus squared of the wavefunction with a rectangular distribution, as was done in 
    (4) above, we use the variance, which gives {$\sqrt{\langle x_i^2 \rangle} = \hbar/(\sqrt{2}mc)$.} Again, this result applies to the model using the reduced mass. To obtain the correct spatial extension for the full two-body system, {we  divide by two per dimension}. This leads to the corresponding volume per virtual pair:  
    \begin{equation}
    V_{\mathrm{pair}} = \frac{1}{2^{\frac{3}{2}}} \frac{\hbar^3}{m^3c_{\mathrm{rel}}^3}{.}
    \end{equation}
\end{enumerate}

With the exception of approach (2), all these methods yield a volume roughly on the order of the Compton wavelength cubed. This consistency is reassuring, as it suggests a physically meaningful scale for the volume associated with virtual particle-antiparticle pairs.

Estimate (1) was employed by Bulanov et al.~\cite{Bulanov:2010aa} to approximate the threshold laser intensity at which {the} vacuum breakdown--i. e., the onset of pair creation processes in vacuum--can be expected to occur.

Now, everything is in place to determine the proportionality factor in front of the electric field in the expression Eq.~\eqref{Displacement} describing the polarization of the vacuum: $\varepsilon_0$, or more precisely, $\varepsilon_0 \chi_{\mathrm{vac}} (0)$. With the quantum harmonic oscillator model and describing the electron-positron pair with the reduced mass, we find the induced electric dipole moment to be \cite{Leuchs:2023aa} 
\begin{eqnarray} 
\langle d_{\mathrm{el}} \rangle  = \frac{e^2\hbar^2}{2m^3c_{\mathrm{rel}}^4} . 
\label{electron positron dipole} 
\end{eqnarray}
The dipole moment induced in {more massive} particle-antiparticle pairs will be smaller, {scaling inversely with the  cube of their mass}. {Assuming that the volume associated with such a pair is proportional to the cube of its Compton wavelength, as discussed earlier, the resulting contribution to the vacuum dipole moment density, $\Delta P_{\mathrm{vac}} = \langle d_i \rangle/{V_i}$,  becomes mass-independent to leading order. This implies that virtual pairs of different masses contribute to the polarization of the vacuum at comparable magnitudes. Consequently, one must sum over all such species, leading to an expression that bears a formal resemblance to the standard QED result for vacuum polarization:}  \footnote{When using Eq. \eqref{alphachivac} to calculate $\varepsilon_0$ from the QED expression Eq. \eqref{alpha large k2}.}
\begin{eqnarray} 
\varepsilon_0 (0)= \frac{e^2}{2\hbar c_{\mathrm{rel}}} \sum_{i}^{\mathrm{e. \, p.}} \frac{q_i^2}{e^2}\frac{\hbar^3}{m_i^3c_{\mathrm{rel}}^3V_i}
\label{epsilonZeroHOM} 
\end{eqnarray}
The electric charges and masses of the elementary particles are denoted by $q_i$ and $m_i$, and $V_i$ is the volume occupied by the particle type $'i'$. For details see \cite{Leuchs:2023aa}. There, with a similar line of arguments, the other parameter in Maxwell’s equations related to the magnetic susceptibility, $\mu_0$ was also calculated:
\begin{eqnarray} 
\frac{1}{\mu_0 (0)} = \frac{e^2 c_{\mathrm{rel}}}{2\hbar} \sum_{i}^{\mathrm{e. \, p.}}  \frac{q_i^2}{e^2}  \frac{\hbar^3}{m_i^3 c_{\mathrm{rel}}^3 V_i} 
\label{muZeroHOM} 
\end{eqnarray}
This allows us to check whether the speed of light comes out correctly. And it does so perfectly, independent of the choice of volume per pair and of how many different types of elementary particles contribute!

The most plausible volume per pair in the list above are the ones referred to under (4) and (5). Inserting the corresponding expressions for all the different elementary particle types $'i'$ into Eq. \eqref{epsilonZeroHOM}, and multiplying $\varepsilon_0$ by {$4\pi \hbar c_{\mathrm{rel}}/{e^2}$} we get an expression for the inverse of the coupling constant:
\begin{eqnarray} 
\alpha^{-1} (0)= 2\pi N \sum_{i}^{\mathrm{e. p.}} \frac{q_i^2}{e^2}
\label{alphaHOM} 
\end{eqnarray}
With $N=(4/\pi)^{\frac{3}{2}}$ or $N=2^\frac{3}{2}$, depending on whether we choose volume option (4) or option (5). There are two unknowns, the number of particle types and the volume per pair. Assuming that $\frac{\hbar^3}{m_i^3 c_{\mathrm{rel}}^3 V_i}$ is independent on the index, the two unknowns are separate numbers simply related by Eq.~ \eqref{epsilonZeroHOM}. If we know both of them we can calculate $\varepsilon_0$ and thus $\alpha (0)$. If we only know one of them, we can calculate the other one using the experimentally determined value of $\varepsilon_0$ (or equivalently $\alpha (0)$). So we can say that Maxwell --assuming that we know the volume, e. g. by using volume options (4) and (5)-- allows us to determine the sum over the squared charges normalized to $e^2$. Solving Eq. \eqref{alphaHOM} for the sum one finds:
\begin{eqnarray} 
\sum_{i}^{\mathrm{e. \, p.}} \frac{q_i^2}{e^2} = 11.5 \pm 3.8
\label{sumq2} 
\end{eqnarray}
The uncertainty for that number 11.5 represents the spread of the volume per pair between options (4) and (5). Evaluating the sum with the particles from the standard model gives $'9'$ in surprisingly good agreement within error bars, considering the simplicity of the harmonic model used here. If one would find a way to determine the volume occupied per virtual pair more precisely, one could further reduce the spread in Eq.~\eqref{sumq2}. Apparently, starting from essentially first principles, the interplay between Maxwell's equations and the standard model of particle physics carries quite far.

The remarkable result presented above establishes a connection between the QED coupling constant and the properties of the vacuum--without encountering divergences. This stands in stark contrast to standard QED. While one might argue that we effectively introduce a relativistic cutoff, the approach goes beyond that. The quantum harmonic oscillator model leaves little room for arbitrary adjustments.

Unlike a crystalline dielectric, which has a well-defined unit cell due to its periodic structure, this model determines the volume occupied by a virtual particle-antiparticle pair through the ground-state wavefunction, imposing strict constraints. Perhaps it is this finite minimal spatial structure that helps to circumvent the usual divergences. Along similar lines, Fried and Gabellini~ \cite{Fried:2012aa} suggested that confined wavefunctions might provide a means to avoid divergences, much like in our simplified model.

Looking back at \eqref{epsilonZeroHOM}, there are two uncertain quantities: First, the number of different types of elementary particles and their charges. Clearly, the sum over the squared charges should be equal or larger than what we can calculate based on the standard model. The second quantity is the ratio of the volume effectively occupied by a virtual pair and the particle's Compton wavelength cubed, assuming this ratio is the same for all particles, which seems reasonable. So, knowing the volume ratio precisely would provide us with information about elementary particles not yet discovered and vice versa.

Some scientists even speculate that spacetime may not be fundamentally continuous or that entirely new physics at extremely high energies could be at play~\cite{Wilczek:2008aa}. In this sense, we find ourselves in a situation similar to that of Planck in 1912--able to formulate ideas but left largely to speculation~\cite{Rafelski:1985aa,Podolny:1986aa}.

\section{\label{model continued} Further {insights into} the vacuum using this simple model}

{Considering} the vacuum as a polarizable medium, {where  particle-antiparticle pairs act as} the polarizable entities,  has also been used to derive other properties of the vacuum. As {originally noted} by Sauter~\cite{Sauter1931} and later {rigorously formalized} by Schwinger~\cite{Schwinger1951}, the vacuum will break down for a { sufficiently strong electromagnetic field, leading to the spontaneous creation of an electron–positron plasma. In this regime, electron–positron pairs} dominate, {to their comparatively small rest mass among all elementary particles.}The simple model can be used to estimated the value for the constant electric field $E_S$ at which the probability for pair creation approaches unity.

Interestingly the electric dipole moment induced by a constant external field calculated using the quantum harmonic oscillator model and the classical harmonic oscillator model yield the same results \cite{AllenEberly}. The force, separating virtual electrons and positrons is given by {  $eE=m\omega_0^2 x$}.  {As previously discussed,} the spatial extension of the virtual pairs should be of the order of the Compton wavelength. Thus, if the induced relative {displacement} $x$ approaches twice the Compton wavelength $x=\hbar/(mc)$ the particle wave functions hardly overlap anymore, a pair is created and the vacuum breaks down. Under this condition, with $\omega_0=2mc^2/\hbar$, as above, 
we obtain for the limiting electric field $E_S$:
\begin{equation}
    E_S= \frac{4m^2c^3}{e \hbar} \, .
    \label{Schwinger-field}
\end{equation}

In \cite{Sauter1931}, Fritz Sauter thanks Werner Heisenberg for {directing his attention to a remark made orally by Niels Bohr, concerning an unusually high electron transmission through a potential barrier exceeding the electron’s kinetic energy. A translation from the original German text in \cite{Sauter1931} reads:} \textit{N. Bohr voiced the presumption, that a transmission can be expected only if the potential step is so steep, that the potential grows over the distance of the Compton wavelength by an amount equal to the rest energy of the electron}.  This gives the same estimate for the limiting field as the one in Eq.~\eqref{Schwinger-field} apart from a factor of two. 

Experiments {aimed at observing} pair creation at the focus of a high-power femtosecond laser are being planned by the European Light Infrastructure (ELI). Currently, the maximum {achievable} intensity {remains} about seven orders of magnitude {below the critical field} $E_S$. {Nevertheless}, as Bulanov et al. \cite{Bulanov:2010aa} pointed out,  {reaching such a high field--where the pair creation probability approaches unity within a single interaction volume--is not strictly necessary. Pair production can occur}  in many ``elementary cells" of the vacuum, {each with a characteristic volume of the order of} Compton wavelength cubed. {It is sufficient for the total pair creation probability, obtained by summing over all such elementary ``cells" within the laser focal volume, to approach unity}. In this way, several orders of magnitude are gained and pair creation in the focus of a laser seems within reach. 

There have even been speculations, whether there might be a further lowering of the threshold by some nonresonant field enhancement when focusing into a nonlinear medium such as the vacuum~\cite{Leuchs2022}.

As mentioned above, the speed of light is determined by vacuum polarization. As Robert Dicke put it \cite{Dicke:1957aa}: \textit{The velocity of light in a 'bare' space could be greatly different from $c$ or even meaningless. ... It is $c$ only after including vacuum polarization effects}.  
Obviously one cannot switch of the polarization of the vacuum and perform experiments in bare space. But it is possible to choose a material with counteracts the polarization of the vacuum~\cite{Veselago:1968aa}, at least in some frequency window, so called epsilon-near-zero or ENZ material; see e.g. \cite{Smith2008,Engheta2008}, or so called negative refraction (for early experimental work see \cite{Russell:1984aa}). Maybe some fundamental test might become feasible with this technology.

{It is remarkable how many fundamental questions can arise from considerations based on the quantum harmonic oscillator model.}

\section{\label{remarks}Discussion and outlook}

More broadly, the vacuum represents the ground state of our universe and the basis of all of our existence. Fields and particles are excited states of the vacuum. Therefore, very understandably, we thrive to understand the vacuum as best as we can~\cite{Lee:1988aa}. Our models and the way we visualize phenomena in nature are a joint effort of the world-wide scientific community to which so many contributed over the years. To make progress, it is good scientific practice to question our prior thoughts and focus on the experimental evidence sharpening our thinking \cite{Englert:2013}. The statement in many undergraduate classes was that the QED vacuum can be polarized but only by charges, there is no linear response to fields. Looking back, this is confusing. In view of the above one would rather say there is a linear response but it is already included in Maxwell's equations. After all, each and every electric field is ultimately created by electrical charges--why should some electric fields polarize the vacuum and others not? Now we have an answer: all electric fields polarize the vacuum, but this vacuum polarization is partially accounted for implicitly in Maxwell's equations and partially through explicit QED diagrams.

In the main text above, one obvious question is, why does the simple harmonic oscillator model for the low-energy response of the vacuum give an answer without any divergence, other than the full QED theory that cannot do without divergencies? We can only speculate. Maybe it is because the harmonic model focuses and is applicable only to the regime of small deviations from the equilibrium of the vacuum ground state. 
There are two different types of vacuum polarizations: an intrinsic one, originating in the position uncertainty of the ground state and another one which relates to induced dipole moments. For the former, the vacuum is in the ground state, the mean polarization is zero and only its variance is nonzero. In contrast, the latter type of vacuum polarization has a nonzero mean value, which is generated by a small deviation from the ground state. This corresponds to a small admixture of the excited state; e.g., of electron and positron, for which we have, however, Pauli's exclusion principle. As a result, the density of induced dipoles cannot be arbitrarily large but is of the order of the inverse of the volume occupied by the three dimensional harmonic oscillator; that is, roughly one over the Compton wavelength cubed. 

Another thought might be that the state of the electron-positron pair should be a singlet state with zero total angular momentum~\cite{Wheeler:1946aa}. According to this argument, we would deal with bosons and the density could be infinitely high. The experimentally determined $\varepsilon_0 \chi_{\mathrm{el,vac}} = \varepsilon_0$  in combination with our harmonic model suggests that electron and %proton 
positron behave independently as fermions near the ground state of the vacuum. This seems to be the reason for avoiding divergencies in our simple model \footnote{Note, that in physics we know many other cases in which such a simple harmonic model can be quite successful, such as for modeling the static polarizability of alkaline atoms.}. 

The model parameters, which we used, are supported by another completely different argument indicating that each virtual pair occupies a finite volume. This is the screening of a bare {pointlike} electric charge by vacuum polarization. This screening only becomes effective at a radius of the Compton wavelength and longer! \cite{Gottfried:1986aa,Leuchs:2020aa}. A reduction of screening in high momentum transfer collisions was observed experimentally as discussed above \cite{Levine:1997,Abbiendi:2004aa}.

The special properties of the vacuum is also relevant in diffraction: Huygens' elementary waves in vacuum remarkably do not scatter backwards (other than in a dielectric), which can also be viewed as a consequence of the properties of the vacuum: It has comparable dielectric and diamagnetic properties. Consequently both, an electric and a magnetic dipole are induced in the vacuum at one point in space, interfering constructively in the forward and destructively in the backward direction. This remarkable property of the vacuum provided guidance in designing metamaterials with similar response~\cite{Geffrin:2012,Nechayev:2019}. Note that in the Huygens-Kirchhoff theory of diffraction, the {backscattering} takes place locally but is suppressed by interference between elementary waves created along the path of propagation \cite{BornWolf:1959}. Thus, our model of describing low-energy properties of the vacuum complies favorably with a number of experimental observations.

There is one other aspect we think one should be careful with. This aspect is the usage of statistical language to describe the quantum world; e. g., ``spontaneous emission", ``mean lifetime", and ``zero- point fluctuations". One has to be cautious not to take this historically understandable statistical language too seriously.

A simple example from the laser laboratory is the following: if one measures the power of a continuous-wave laser with a {photo diode} and records it as a function of time, then one will readily get a time series which seems to fluctuate. A Fourier analysis of this time series shows high noise amplitudes at low frequencies. This noise rolls off as the noise frequencies increases until it levels off at values not higher than a few MHz, depending on the type of laser. From there on towards even {higher frequencies} the noise is frequency independent, so-called white noise. In all laser laboratories such measurements are done and this white noise is readily seen if the detector is sensitive enough and has low enough intrinsic electronic  noise. This white noise is called quantum noise and it is said to extend to infinity. What does that mean? A coherent state; {i.e., the state of a single-mode laser} without technical noise, when viewed in phase space using Wigner distribution function, is merely a displaced vacuum state. So measuring the quantum noise of the laser is in a way measuring the Wigner distribution of the vacuum state \footnote{A more sophisticated way of measuring the Wigner distribution of the vacuum was demonstrated by Lvovsky et al. \cite{Lvovsky:2001} using a balanced homodyne detector.}. 

Is this a proof for the concept of statistical zero-point fluctuations? The answer is no! Because, if the light intensity and thus the associated electric field were really oscillating up to infinitely fast, then according to Maxwell's equations one should find infinitely high values of $\dot{\vec E}$ and thus of $\vec{\nabla}\times\vec B$. But we do not observe such infinitely large terms in the lab. The explanation is that quantum wave functions are not intrinsically ``noisy", they are not subject to statistical fluctuations. According to quantum physics, the vacuum has {zero-point} uncertainty and so has the electric field of the laser. Only when we measure the intensity or the field --with a photodiode or a homodyne detector, respectively-- then something weird happens! In the quantum measurement process the measured quantity is projected on to one of the --in general, several-- possible values. When preparing and measuring a quantum system repeatedly in an identical way, the quantum uncertainty is transformed into an apparent fluctuation \cite{Korolkova:2024}.  In the case of a laser in a coherent state, this results in Poissonian photon number fluctuations. Such noise, which appears only through the measurement, is called projection noise \cite{Wineland:1994}. This exemplifies that one has to be careful to distinguish what is uncertain and what is fluctuating. Before any measurement a quantum system typically undergoes a unitary evolution and thus a deterministic evolution. 
%This is by the way the reason why a quantum computer is conceivable. 
It is the measurement which introduces statistics --- 
But strict quantum correlations survive the measurement which is why a quantum computer is conceivable and can be expected to show enhanced performance for special types of problems. The way we discuss the quantum dynamics above, we stay close to an interpretation of quantum physics, which assumes that the wave function describes a quantum wave, which really exists. Other interpretation discuss the scenario differently \cite{leuchs2015}, but so far there is no way to decide in favor of one of these interpretations based on experimental results.

All experimental evidence we have is that the projection associated with a quantum measurement, happens as fast as the detector allows for. The projection seems to have no intrinsic time scale. Therefore, physicists abandoned the older name ``collapse of the wave function", an expression that seems to suggest some dynamical evolution. The more abstract notion of a mathematical projection is widely preferred these days. But who knows, maybe one day when we can measure faster and faster, we may find an intrinsic time scale associated with the measurement projection. Rudolf Haag hypothesized in this direction trying to inspire new experiments~\cite{Haag:1990,Haag:2013,Englert:2013}. 
Maybe there is even a grid in all four dimensions of space-time and a fundamental smallest time step is the answer to how fast a projection can happen. So far it happens as fast as we measure and anything else is speculation. 

In any case, visualizing the vacuum as consisting of fluctuating virtual particle-antiparticle pairs should be handled with care in view of the above. A better way is to view the vacuum as the ground state with respect to different particle types and fields with stationary uncertainty, much in the same way one would deal with other quantum systems.

In conclusion, it is remarkable and surprising that our attempt to calculate $\varepsilon_0 \chi_{\mathrm{el,vac}} = \varepsilon_0$ 
%as experimentally determined from measurement in classical optics and electromagnetism 
based on the properties of the quantum vacuum using the harmonic model matches fairly well with the experimental value as determined from measurement in classical optics and electromagnetism; and that using \eqref{alphachivac} the model likewise yields a value for the fine structure constant $\alpha$ for $k^2=0$ which is in the right ball park.
%perfectly with  as determined experimentally from high energy particle collisions using \eqref{alphachivac}.

\begin{acknowledgments}
We acknowledge helpful suggestions by the late Joseph H. Eberly of Rochester, as well as by the late Nikolay Borisovich Narozhny of Moscow. Likewise we very much enjoyed our close collaboration with Margaret Hawton of Thunder Bay. We also acknowledge discussions with Kimball A. Milton of Oklahoma and with Michael Thies of Erlangen, who conveyed both criticism and insight. Last not least, we appreciate helpful suggestions by Anatoly Mikhailovich Kamchatnov of Troitsk.
\end{acknowledgments}

%\bibliography{Uspekhi}


\end{document}